\documentclass[usenatbib]{mn2e}

\usepackage[dvips]{graphicx}
\begin{document}

\title[Low-metallicity stellar haloes]{
Low-metallicity stellar halo populations as tracers of  dark matter haloes}

\author[Tissera et al. ]{Patricia B. Tissera$^{1,2,3}$,
Cecilia Scannapieco$^{3}$\\
$^1$ Departamento de Ciencias Fisicas, Universidad Andres Bello, Av. Republica 220, Santiago, Chile.\\
$^2$ Instituto de Astronom\'{\i}a y F\'{\i}sica del Espacio,
CONICET-UBA, Casilla de Correos 67, Suc. 28, 1428, Buenos Aires,
Argentina.\\
$^3$Millenium Institute of Astrophysics (MAS), Santiago, Chile.\\
$^4$Leibniz-Institut f\"ur Astrophysik Potsdam (AIP), An der Sternwarte 16, D-14482 Potsdam, Germany.
}

\maketitle

\begin{abstract}
We analyse the density profiles of  the stellar halo populations in eight Milky-Way mass galaxies,
simulated within the $\Lambda$-Cold Dark Matter scenario. We find that
accreted stars can be well-fitted by an  Einasto profile,
as well as any subsample defined according to metallicity. 
We detect a clear  correlation between the
Einasto fitting parameters of the low-metallicity stellar populations and those of the dark matter haloes. The 
correlations for stars with [Fe/H]$<-3$ allow us to predict the shape
of the dark matter
profiles within residuals of $\sim 10 $ per cent, in case the
contribution from in situ stars remains small. Using Einasto parameters
estimated for the stellar halo of the  Milky Way and assuming the later formed with significant contributions from accreted
low-mass satellite, our simulations
predict   $\alpha \sim 0.15 $ and $r_2 \sim 15$ kpc for its dark
matter profile. These values, combined with observed estimations of the
local dark matter density, yield an enclosed dark matter mass at
$\sim 8$
kpc in the range $3.9 - 6.7 \times 10^{10}$ M$_{\odot}$, in agreement with
recent observational results.
These findings suggest that low-metallicity stellar haloes could store
relevant information on the DM haloes. Forthcoming observations
would help us to further constrain our models and predictions.

\end{abstract}

\section{Introduction}

The stellar halo populations are  important components of galaxies
which extend around them out to few hundreds of kiloparsecs. Observations
show that they are complex structures where  debris from stripped satellites, stars, and surviving
satellites can be identified within diffuse stellar envelopes. 
Stellar halo populations
have been detected out to several kiloparsecs from the galaxy centre in  nearby galaxies, regardless of their morphology
\citep[e.g.][]{rejkuba2011,sarajedini2012, crnojevic2013,ibata2014}.
There are also evidences of  stellar haloes in galaxies at $z\sim
1$ \citep{trujillo2012}.
The Milky Way (MW) is the best observed stellar halo where detailed
kinematics and chemical abundances for individual stars are being obtained from different surveys,
allowing the identification of chemical-dynamical patterns
\citep[e.g.][]{caro2007, nissen2010, deason2011, sheffield2012,beers2012,schonrich2014}.
Such a detailed analysis is now being extended to nearby galaxies as
in the case of  M31 and M32 \citep[e.g.][]{sarajedini2012, ibata2014}. These
observational results provide invaluable information to constrain  cosmological models.

Simulations within a hierarchical clustering scenario show that most
of the stars in the stellar haloes come from accreted satellites in
wet or dry mergers \citep[e.g.][]{bullock2005,purcell2009,zolotov2009,cooper2010, font2011, tissera2012, tissera2013,pillepich2014}.
Hydrodynamical simulations, including chemical evolution, produce 
stellar halo populations that are old and high $\alpha$-enriched in
agreement with observations. \citet{tissera2014} show how the
metallicity distribution of stellar haloes vary with the radius
depending on the accretion histories.
Simulations also detect the existence of a fraction of stars formed
{\it in situ}, which could explain the presence of stars
with lower $\alpha$-enhancement and significant rotation velocities
reported in observations of the MW
\citep[e.g.][]{sheffield2012}. 

Stellar haloes contain an important fraction of old and metal-poor
stars, which are a window towards the first epoques of the formation of
galaxies \citep[e.g][and references there in]{placco2014} and hence,
might store information relevant to galaxy formation. 
 Using
 cosmological simulations of MW mass  galaxies,
\citet{tissera2014} reported
 the fraction of low-metallicity stars to increase with
 radius. They also found about $\sim 60-90$ per cent of these stars
formed in accreted low-mass satellites ($M < 10^9$~M$_\odot$). Low-metallicity stars populate
the whole stellar haloes, mapping the potential well more  uniformly
out to the virial radius.

The stellar halo populations inhabit the potential
well determined by the dark matter  (DM) mass. As satellites fall in, they are
disrupted, contributing with stars and gas to the haloes. 
 Unlike the gas that  experiences significant  dissipation
during infall, the stars are expected to store information for a longer
time \citep{eggen1962}. In particular, the low-metallicity
stars might be the best tracers of the potential well, as they come in
small mass satellites which are easily disrupted and are mainly subject to
the gravitational potential well dominated by the DM.

The aim of this paper is to explore if the
stellar halo populations, and particularly the low-metallicity stars, can
trace the  DM distributions.  
For this purpose, we use  a suite of MW mass haloes from the
Aquarius Project which were modified to include gas dynamics.
The chemo-dynamical properties and
history of assembly of the stellar halo populations of
these  haloes have been extensively studied by
\citet{tissera2012,tissera2013,tissera2014}.
The DM profiles of six of them 
were analysed by \citet{tissera2010}.

This letter is organized as follow. Section 2 describes the numerical
simulations and models. In Section 3 we analyse and discuss the density
profiles of stellar halo populations and correlate them with those of  their dark
matter haloes. In the conclusions section,  we summarize our main results.

\section{The simulated haloes}
We analysed  the suite of eight MW mass-galaxy systems
of the Aquarius Project reported by \citet{scan09}, which  were
modified to include baryons.
 The systems  were selected from a cosmological volume of $100 \ { \rm Mpc \  h^{-1}}$ box, 
 with only a mild isolation criterium
imposed  at $z=0$ and the requirement to have similar total mass  to the MW (see \citet{sprin2008} for further details on the 
generation of the initial conditions). 
The simulations are consistent with a  $\Lambda$-CDM cosmogony with  $\Omega_{\rm m}=0.25,\Omega_{\Lambda}=0.75,
\sigma_{8}=0.9, n_{s}=1$ and $H_0 = 100 \ h \ { \rm km s^{-1} Mpc^{-1}}$ with $h =0.73$. 
 The simulated haloes have $\approx  1$ million total particles within the virial radius and 
virial mass in the range $\approx 5 - 11 \times 10^{11} {\rm M_\odot} h^{-1}$. 
Dark matter particles have masses of the order $\approx 10^{6}{\rm M_\odot }h^{-1}$
while initially  gas particles have   $\approx 2 \times 10^{5}{\rm
  M_\odot} h^{-1}$.  More details on the simulated galaxies can be
found in \citet{scan09} and \citet{tissera2012}.
We used a version of  {\small GADGET-3} which
 includes stochastic star formation, a physically-motivated Supernova (SN)
 feedback model and chemical enrichment by Type II and Ia SNe.
 Details on the
 SN model can be found in \citet{scan05,scan06}.
\citet{scan09, scan10,scan11} studied the formation of the disc and the spheroidal components of
these systems and their dynamical and photometric properties. The main
galaxies have  a variety of morphologies with different bulge-to-disc
ratios.

We follow \citet{tissera2012,tissera2013, tissera2014} and apply a criterium based on
angular momentum and binding energy to separate the different dynamical
components. Stellar halo stars are a subset of the spheroid stars,
those for which the binding energy is lower than the minimum one at the half-mass radii of the galaxies. Note that we do not consider in the stellar
haloes particles belonging to friends-of-friends groups of more than
20 particles. Hence, our stellar haloes do not contain satellites, but they do include streams and small substructures from disrupted satellites.

\citet{tissera2013} and \citet{tissera2014}  analysed the formation of
the stellar
haloes and in particular, how to use the chemical
distributions to set constrains on the history of mass assembly. 
In our simulations the stellar diffuse haloes are mainly made by  debris stars
which formed in accreted satellites (outside or inside the virial
radius) and disc-heated stars (so-called {\it in situ}). The latter are found
to be more centrally concentrated.  
\citet{tissera2014} showed  that stellar haloes formed by the
accretion of massive subgalactic systems  have steeper metallicity
gradients.  They also reported that low-metallicity stars are mainly
contributed by low-mass satellites and are more frequent in the
outskirts of haloes.

\section{The relation between the stellar and DM haloes}

We  construct the density profiles of the stellar halo populations by  using
spherical shells  from $\sim 2$ kpc $h^{-1}$ to the virial radius,
centred at the  galaxy centre.  
The density profiles are constructed for the {\it in situ} and
accreted stellar components, separately. Within the accreted stars, we
include debris (i.e. stars formed in satellites before coming into the
virial radius of the main galaxy) and endo-debris (i.e. those stars
formed in gas-rich satellites within the virial radius of the main
galaxy). 
We define two population subsamples  according to their metallicity:
 very metal-poor (VMP) and extremely metal-poor (EMP)
by requiring them to satisfy: [Fe/H]$ <-2$ and  [Fe/H] $<-3$, respectively.

\begin{figure*}
\hspace*{-0.2cm}\resizebox{4.5cm}{!}{\includegraphics{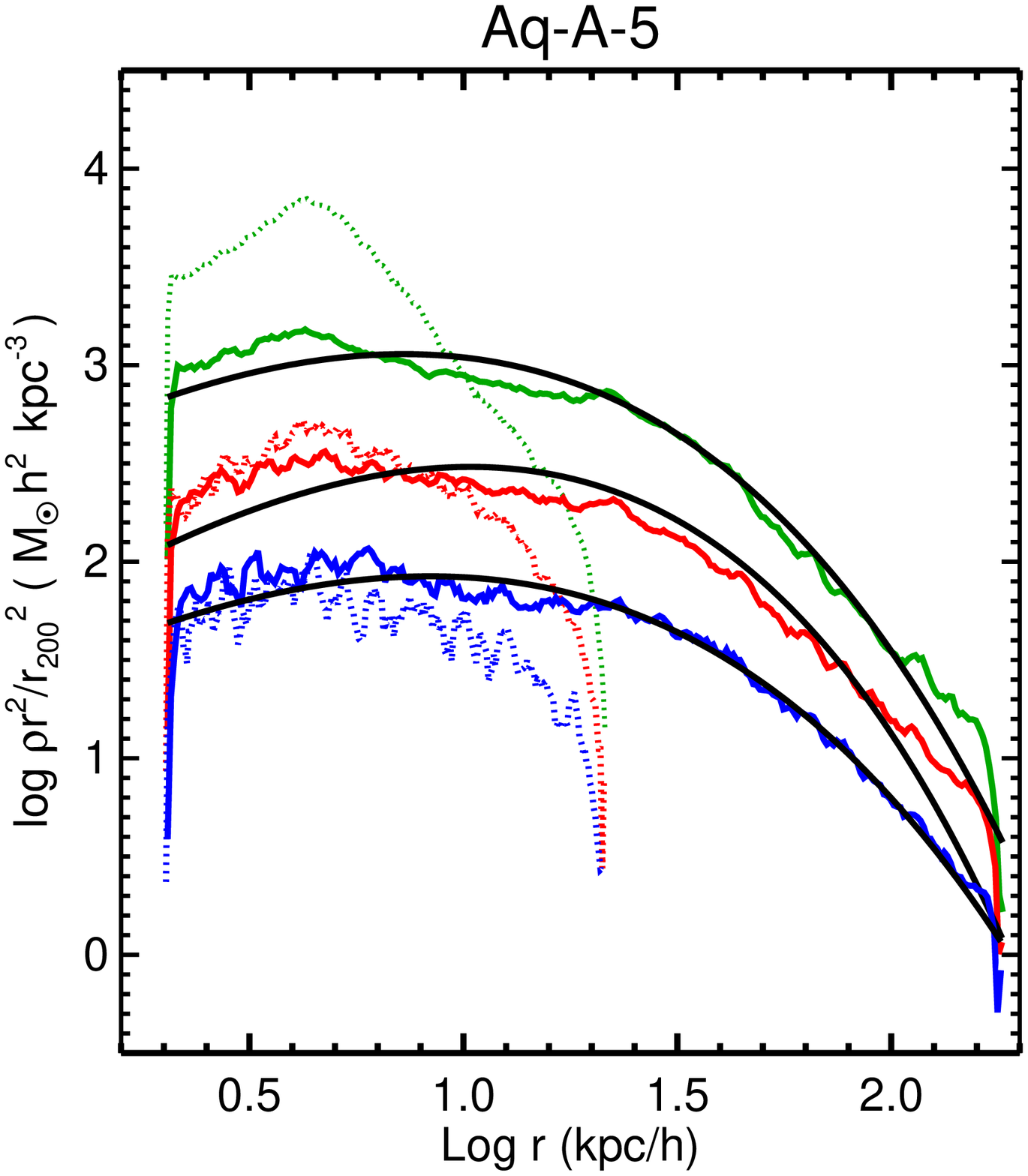}}
\hspace*{-0.2cm}\resizebox{4.5cm}{!}{\includegraphics{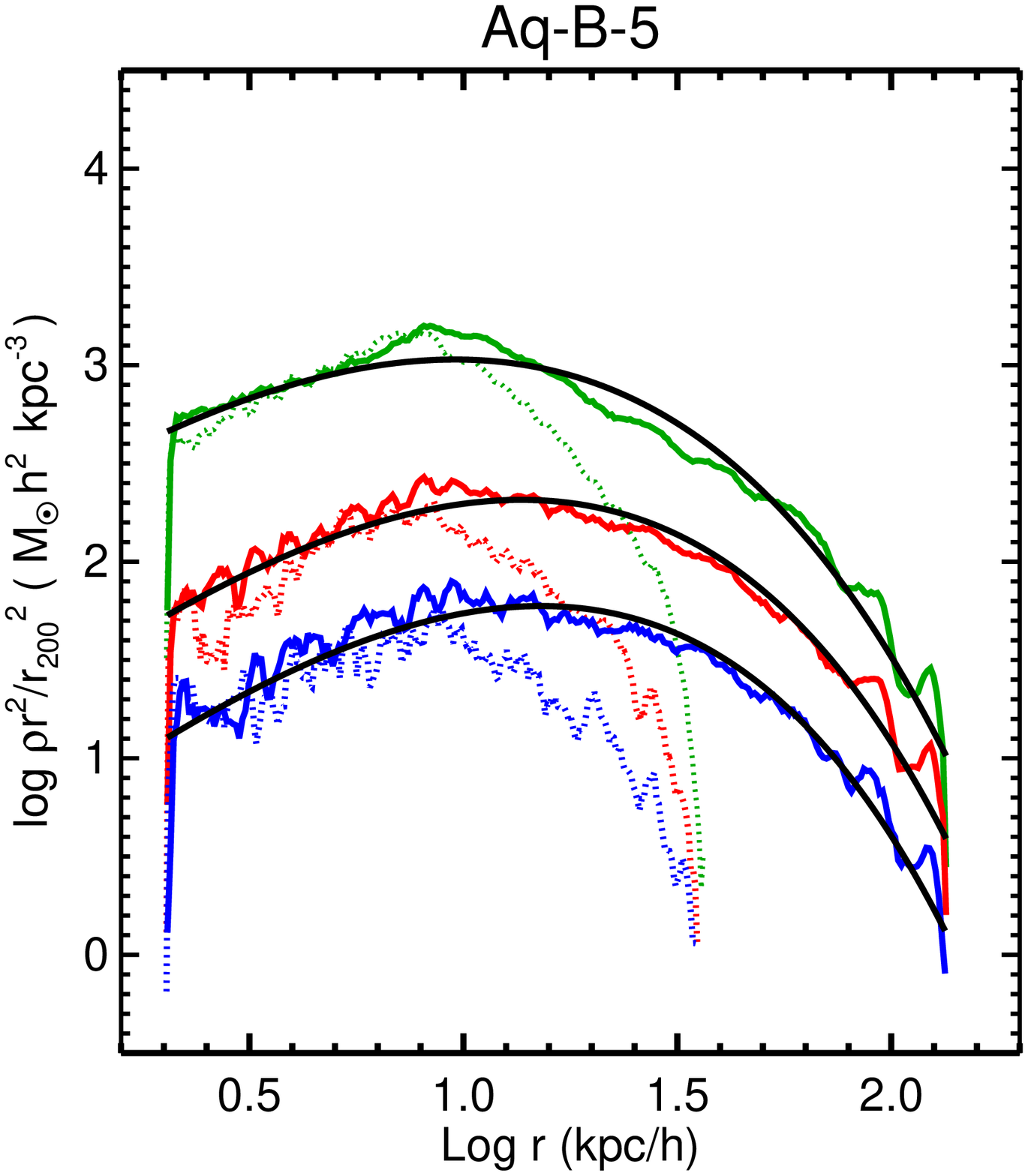}}
\hspace*{-0.2cm}\resizebox{4.5cm}{!}{\includegraphics{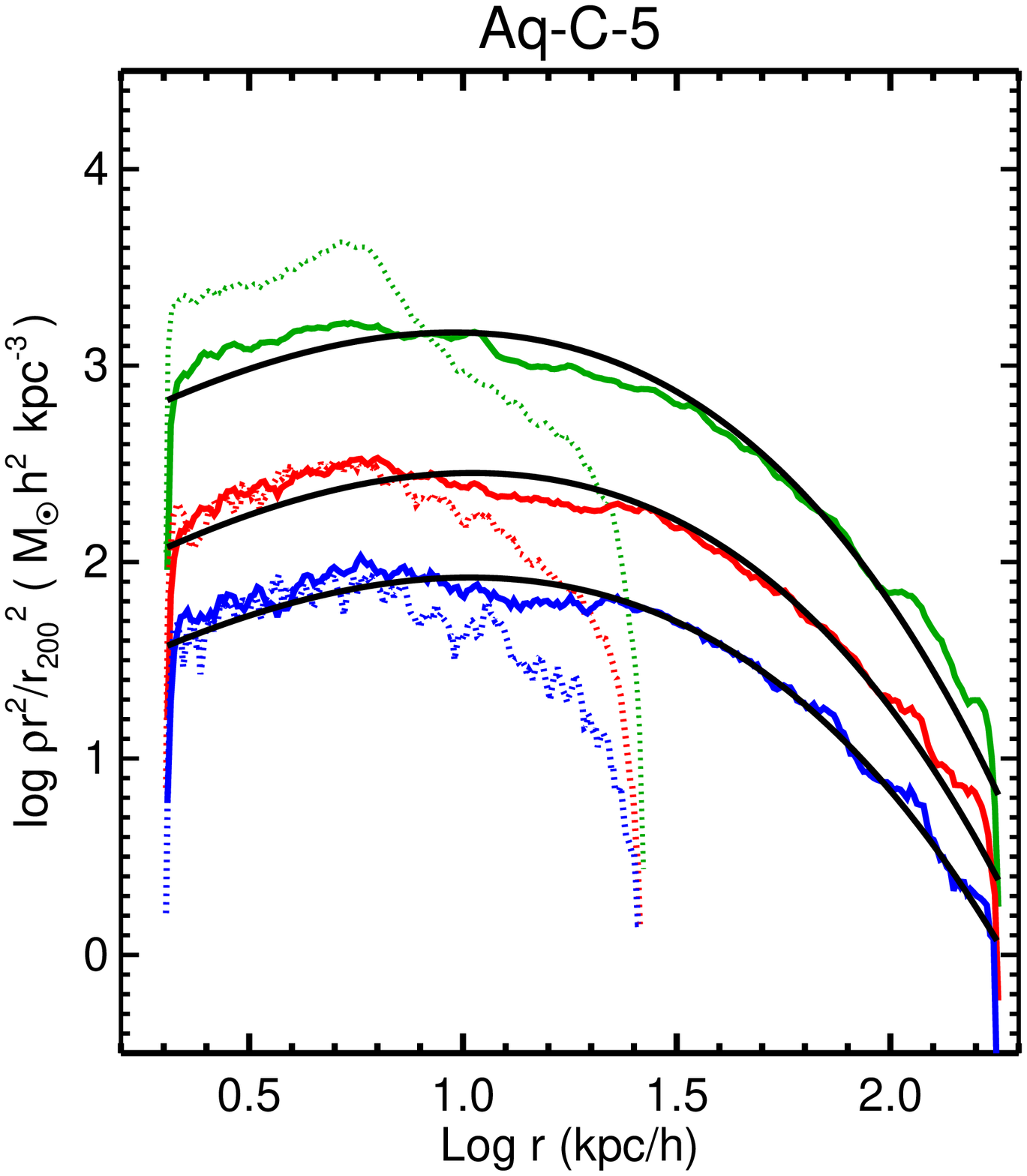}}
\hspace*{-0.2cm}\resizebox{4.5cm}{!}{\includegraphics{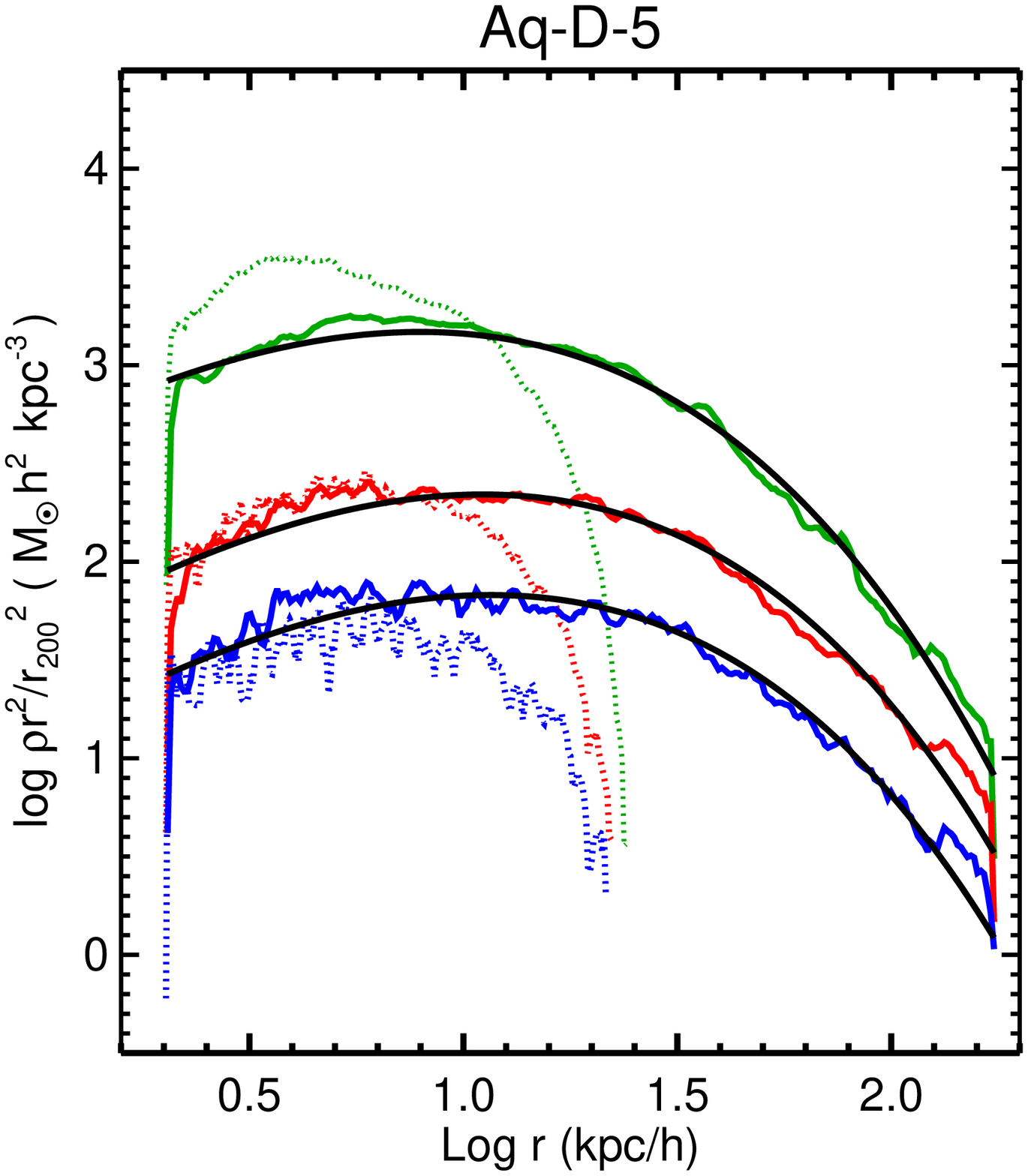}}\\
\hspace*{-0.2cm}\resizebox{4.5cm}{!}{\includegraphics{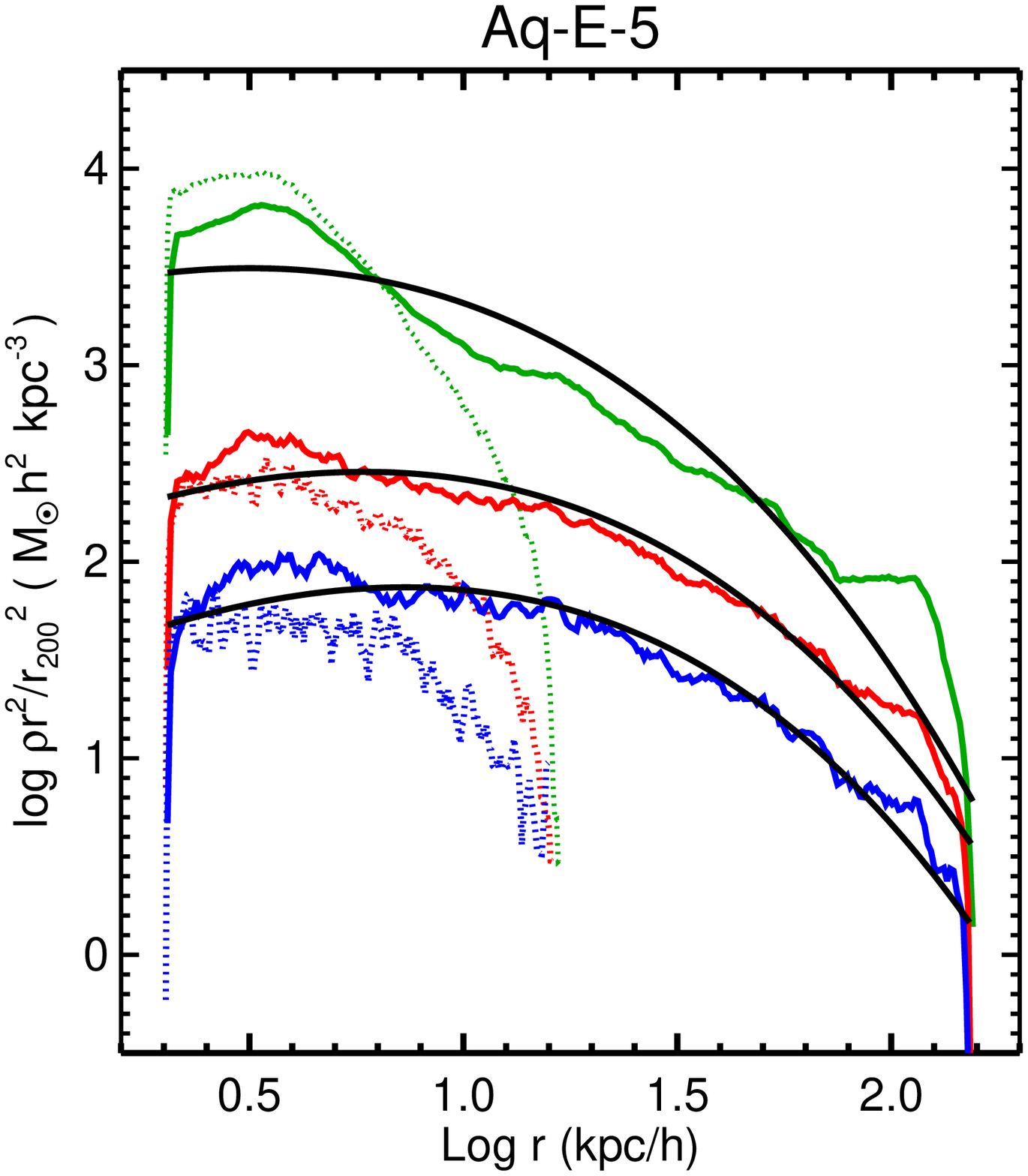}}
\hspace*{-0.2cm}\resizebox{4.5cm}{!}{\includegraphics{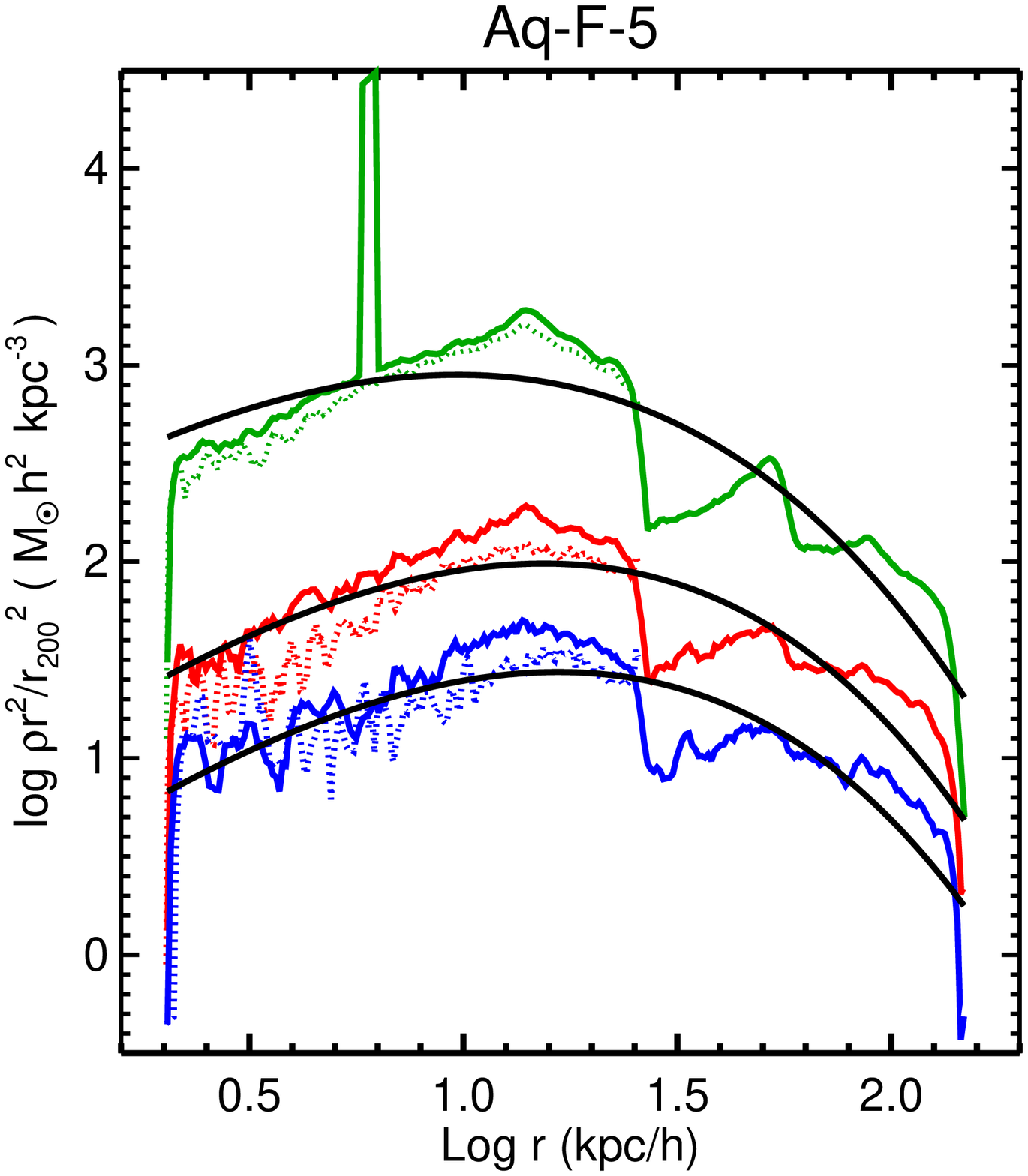}}
\hspace*{-0.2cm}\resizebox{4.5cm}{!}{\includegraphics{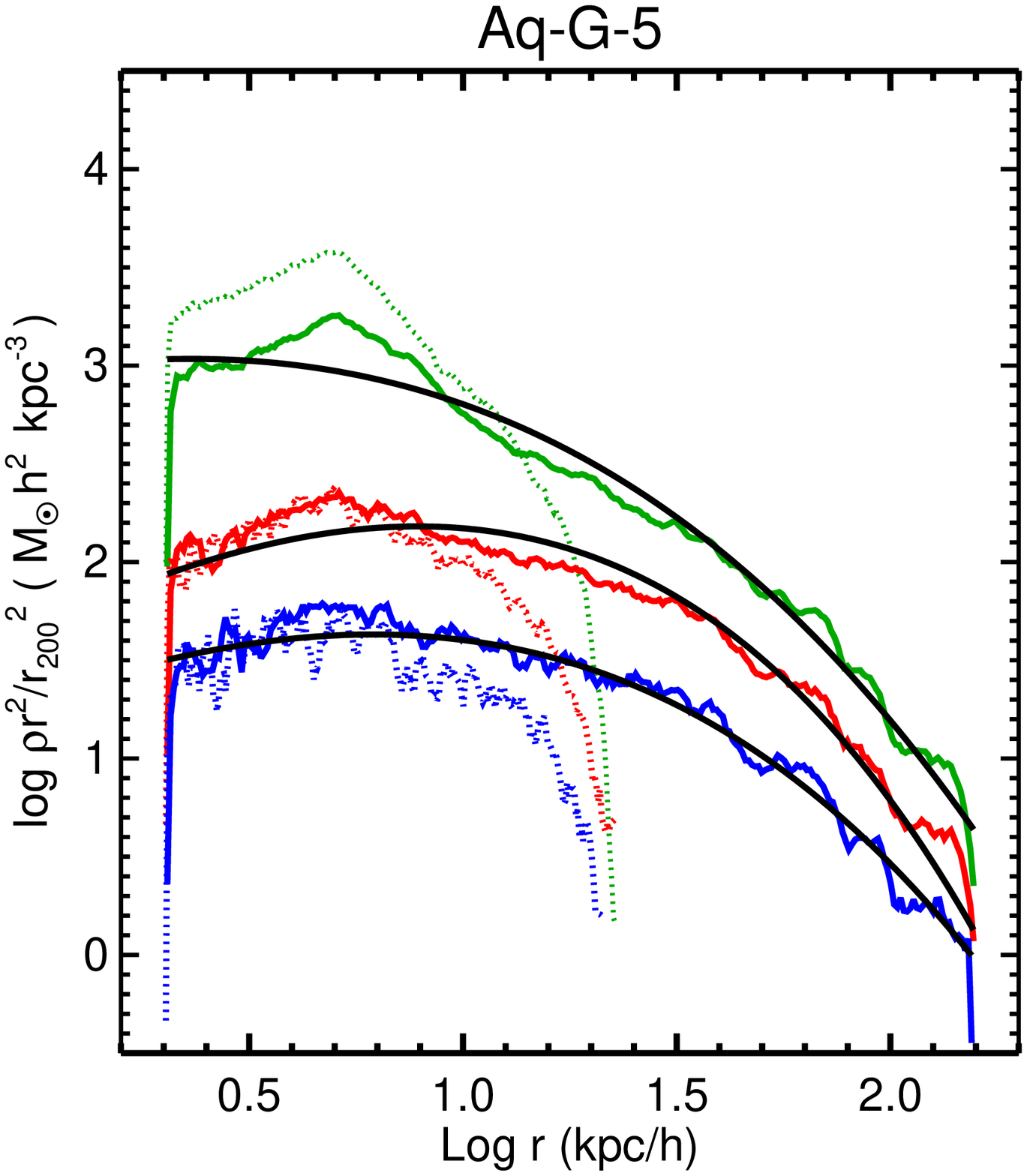}}
\hspace*{-0.2cm}\resizebox{4.5cm}{!}{\includegraphics{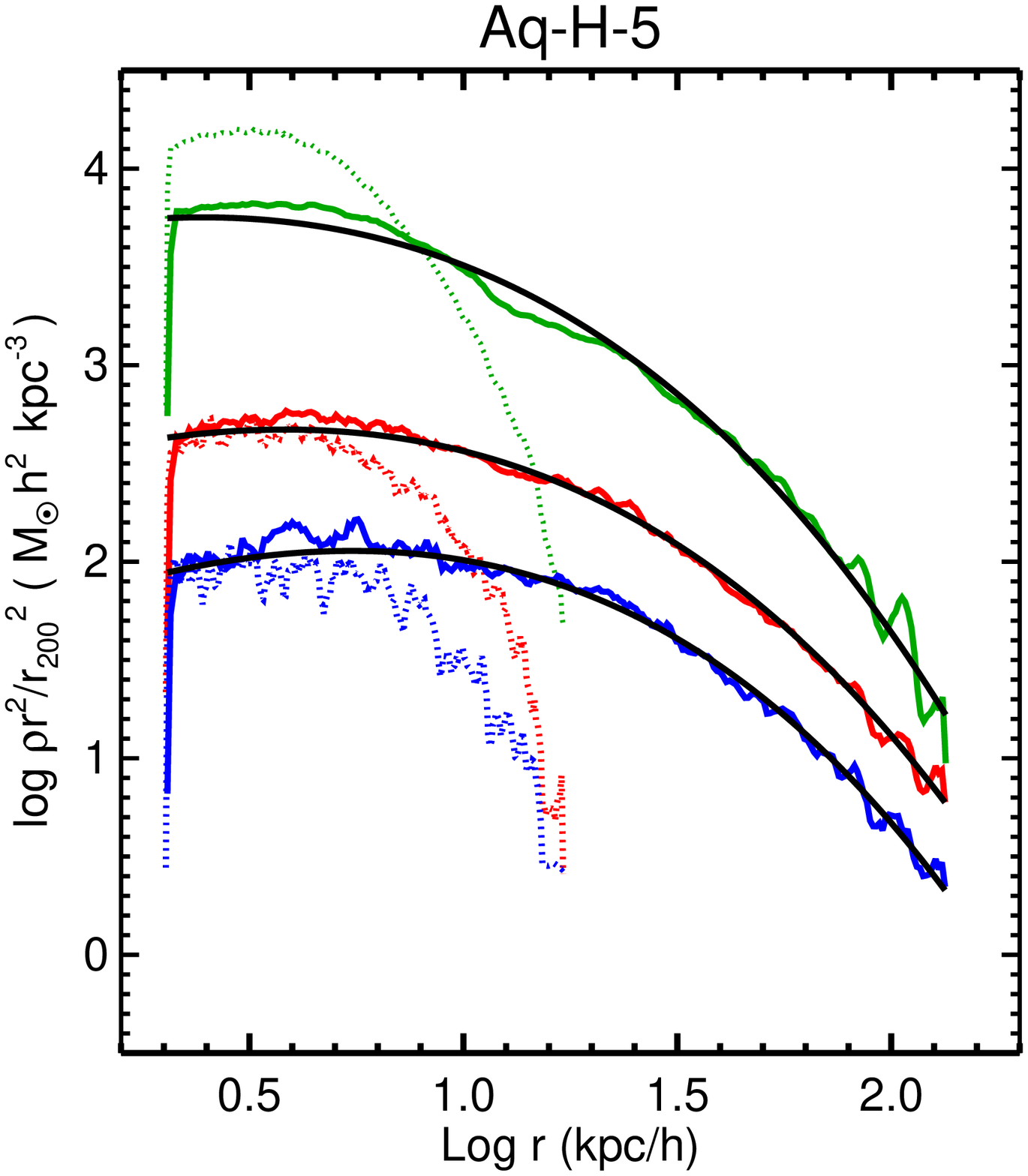}}
\vspace{-0.3cm}
\caption{Density profiles of the {\it in situ} (dashed line)
  and accreted (solid line) stellar
  halo components  (green lines) and the corresponding
  density profiles obtained by applying metallicity thresholds: [Fe/H]
  $ < -2$ (VMP; red) and [Fe/H] $< -3$ (EMP; blue).
The black solid lines depict the Einasto  profile fit for each subsample.}
\label{dmprofiles}
\end{figure*}

We estimate the density profiles for each metallicity subsample  and for
both types of stellar populations.
The profiles for the  {\it in
  situ}   (green dotted lines)
 and
accreted  (solid green lines) stellar components, without metallicity
segregation, are shown in  Fig. ~\ref{dmprofiles}.  These distributions
show how  {\it in situ} stars are more concentrated in the central regions while  accreted ones
extend up to the virial radius. There are two haloes with
larger  relative {\it in situ} contributions: Aq-A-5 and Aq-H-5. In
the case of Aq-B-5 and Aq-F-5, the contributions of  {\it in situ}  are
very small.

 The lower subsequent set of profiles included in
 Fig. ~\ref{dmprofiles} are those obtained 
for the VMP (red lines) and  EMP (violet lines) subsamples.  The contributions from
{\it in situ} stars diminish systematically  as the metallicity threshold
gets more stringent,  since these stars
have a large probability to  have been formed from recycled material, polluted
by previous stellar generations. Disc-heated stars are younger than
accreted stars and have larger $\alpha$-enhancements, on average \citep{tissera2012,tissera2013}.
Hence, as we move to  lower metallicity subsamples, the density
profiles   are mainly determined by 
accreted stars. As discussed by \citet{tissera2014}, accreted stars are
mainly contributed by small galaxy  satellites (few times
$10^{9}$~M$_\odot$) and have a
frequency which increases outwards.

 We fit Einasto profiles to the density distributions from $\sim 2$
 kpc $h^{-1}$ to the virial radius.
The free parameters are  $n=1/\alpha$, $r_{2}$  and $\rho_{2}$ which
indicate the sharpness of the profiles and the radius and density
where their logarithmic slope takes the isothermal value,
respectively.
The accreted stellar haloes are resolved with more than
100000 star particles while these numbers decrease as more stringent
metallicity thresholds are imposed.  The accreted EMPs are resolved by  $\sim
5000-10000$ star particles.

In these simulated haloes, the  density profiles determined by the accreted stars  are  well-fitted by the 
Einasto profile, regardless of the metallicity threshold adopted
(Fig. ~\ref{dmprofiles}; solid black lines). 
This is not the case for the whole stellar haloes because of the
contribution of {\it in  situ}  stars which are more concentrated
than accreted ones. However,  the contribution of {\it in situ} stars
to the EMPs is  between $10-20$ per cent of the total EMP mass. 
And in fact, the Einasto profile
 also yields a good fit for the whole EMPs.

\begin{figure*}
\hspace*{-0.2cm}\resizebox{5.cm}{!}{\includegraphics{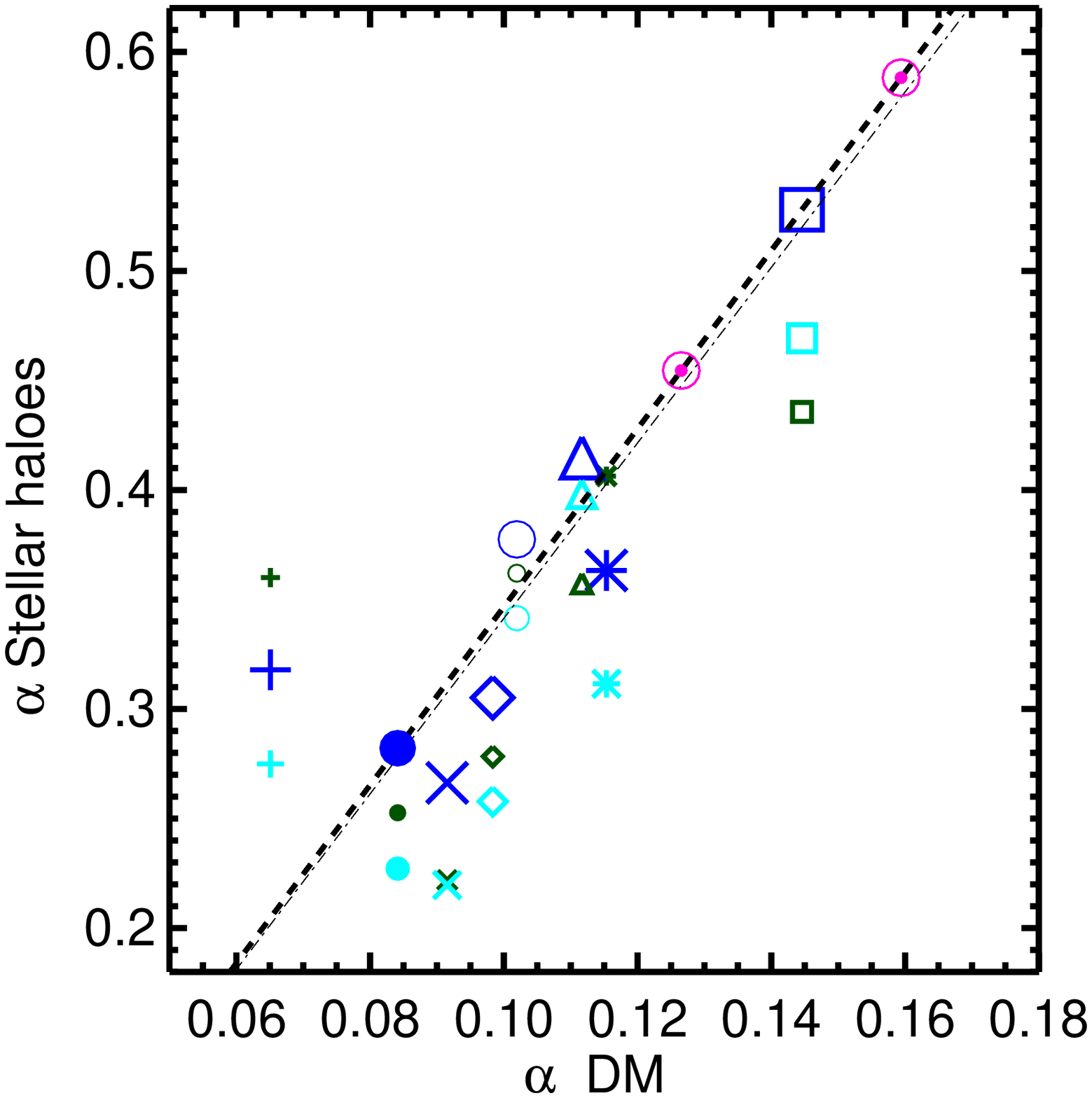}}
\hspace*{-0.2cm}\resizebox{5.cm}{!}{\includegraphics{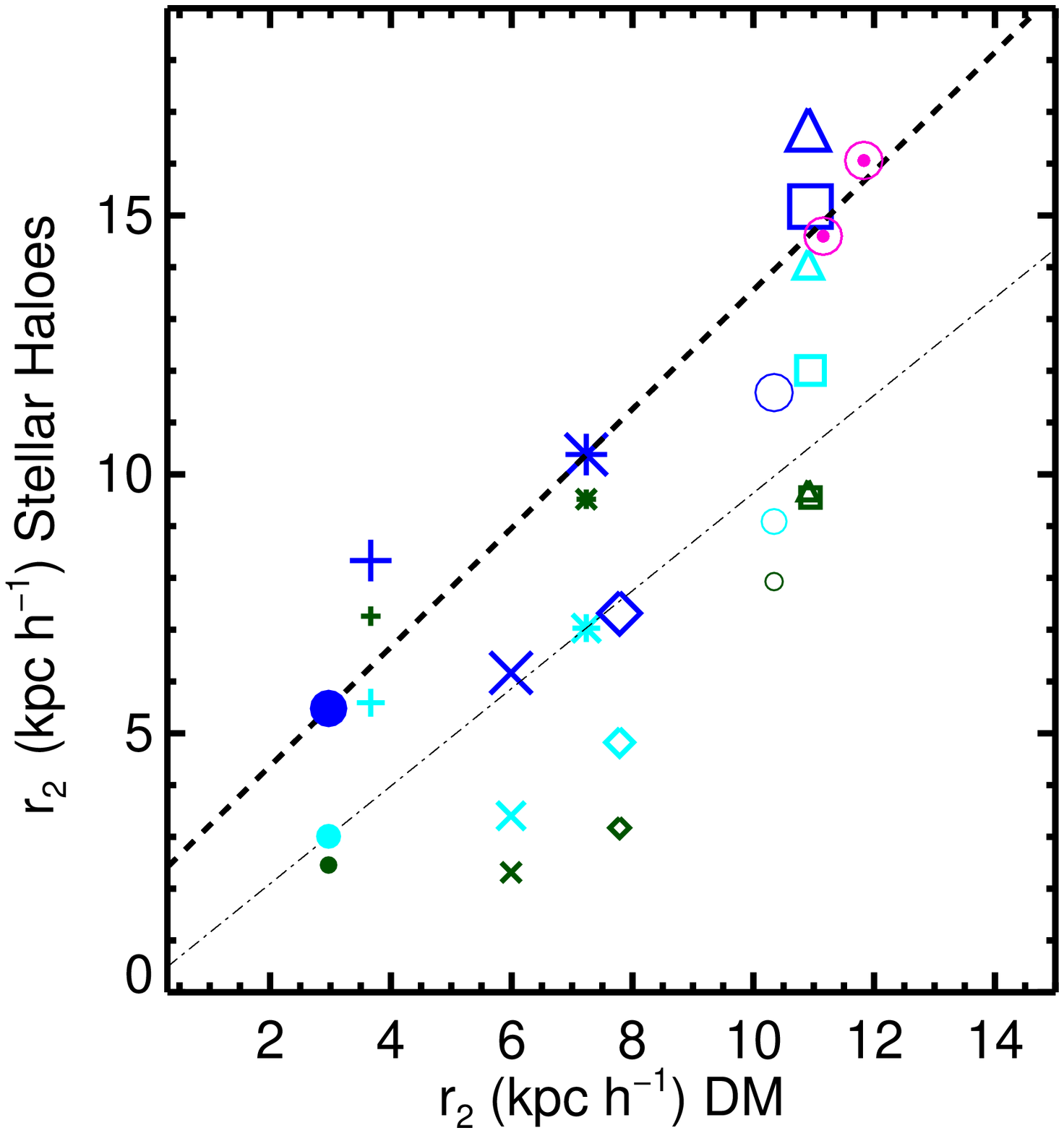}}
\hspace*{-0.2cm}\resizebox{5.cm}{!}{\includegraphics{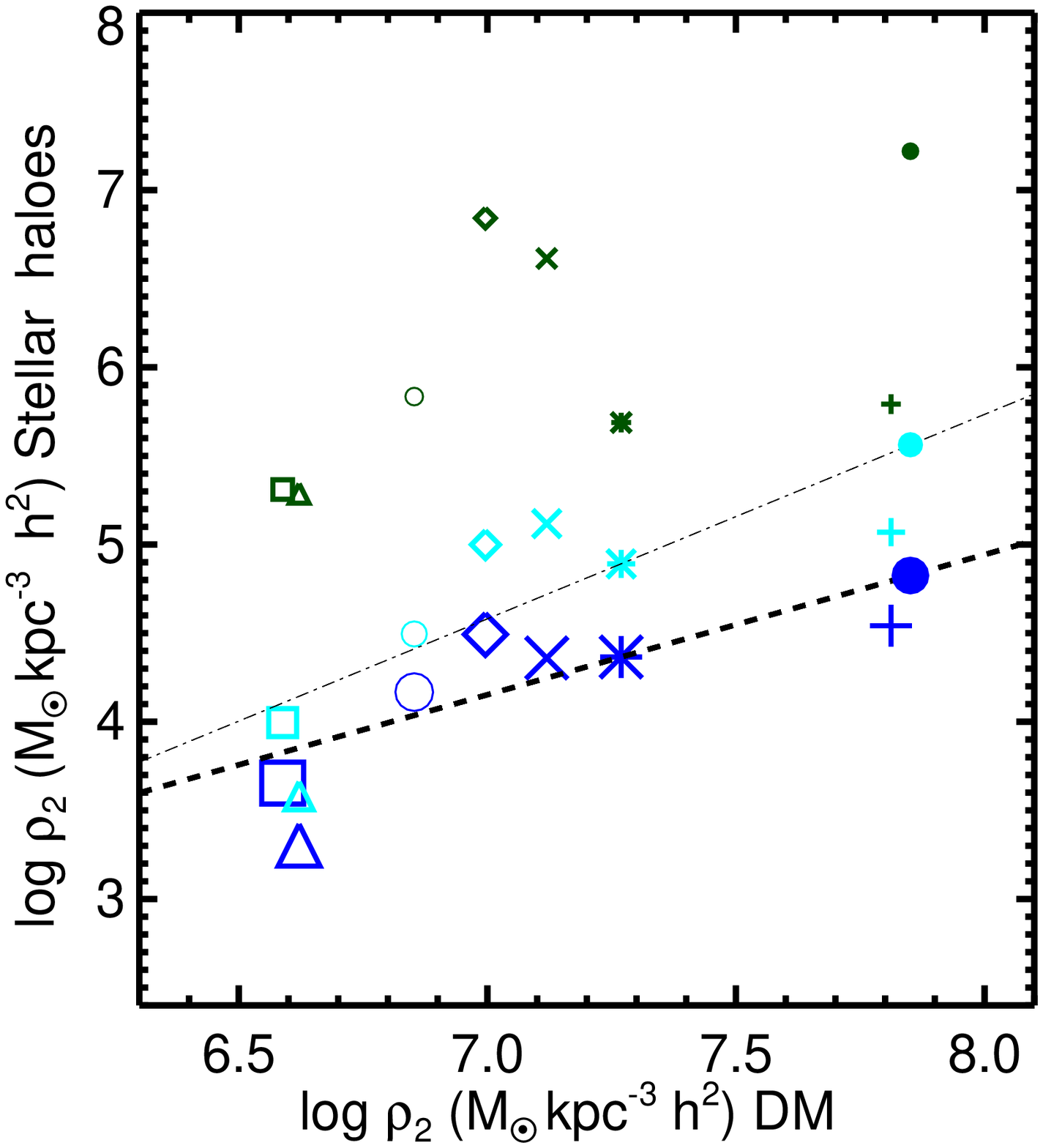}}
\vspace{-0.3cm}
\caption{Einasto fitting parameters  of the density
  profiles  estimated for the  total  accreted stars in the haloes (green symbols), for
 accreted EMPs (blue symbols) and for the whole EMPs
 (cyan symbols) as a function the corresponding best-fitting
parameters obtained for the DM profiles. The fitted linear regressions
to the whole and accreted EMP  parameters are
also shown (think and thick dashed lines). We include predicted $\alpha$ and $r_2$ predicted for the
MW (magenta symbols), using  available observed estimations under the hypothesis that  MW halo formed
mainly by accreted stars from low-mass satellites. 
}
\label{corre_dark}
\end{figure*}

The DM profiles of these haloes were studied by
\citet{tissera2010}, who found that   they are well-fitted by Einasto
profiles (in the same radial range). The final properties of
the DM 
profiles are   the outcome of the joint evolution with their main
galaxy and consequently, they have been already
modified by its formation.
In order to explore at what extent the accreted stellar haloes can
trace the DM ones, we
correlate the Einasto fitting parameters for our metallicity-selected 
subsamples of accreted stars with those obtained for the corresponding DM 
profiles  by  \citet{tissera2010}. 
 Statistically significant correlations for the $\alpha$,  $r_2$ and $\rho_2$
parameters are measured, principally for  
the accreted and whole EMPs, which yield  correlation factors within
the range $ \sim 0.80-0.85$. These factors go down to $\sim 0.50-0.65$  
for all accreted stars and VMPs. Hence, hereafter, we  only analyse the EMPs.
We fit  linear regressions to the 
correlations defined by both EMP subsamples, by applying  a  Levenberg-Marquardt least-squares
algorithm.
In  Fig. ~\ref{corre_dark}, we   show the linear regressions determined
by  the accreted and whole EMPs.
The $\it rms$ found for the $\rho_2$ and $n$ linear fittings are smaller than
$\sim 0.02$ but, for the $r_2$  characteristic scales, it is $\sim
0.2$. For the whole EMP subsample, we also find an increase of a
factor of four in the rms of  $\rho_2$.

We use these correlations to estimate the DM density profiles by
using the fitting parameters of the EMPs and the linear
relations shown in Fig.2. Fig. 3 shows the predicted and simulated
DM profiles, as well as the best-fitting Einasto DM profiles
obtained directly from the simulated data.
Although the shape of the DM profiles can be well estimated from those
of the  EMPs, the normalization is more difficult
to predict without an independent estimation of, e.g., the total
DM mass. For this reason, we renormalized
the profiles so that the predicted and real DM profiles agree
at $r_2$. We could instead use our predicted DM profiles
to infer the DM mass of the systems; in this case the
DM mass is recovered with a $\sim 0.7$ dex dispersion.
This large dispersion reflects the problem with the normalization
of the predicted profiles, indicating the need to have
an independent measure of the DM mass.
The small panels of Fig.3 show the residuals of the DM profiles
obtained by fitting the real data and those predicted by the EMPs to the
simulated DM profiles.
Except for Aq-A-5 and A-H-5, the galaxies with the largest
in-situ fractions, the residuals are always smaller than ten per cent.
These results show that the whole EMP subsample can also trace the shape of the
dark matter profile. This is  because it is dominated by accreted
stars. We also tested the whole and accreted VMPs but in this case,
there are larger dispersions and the predicted profiles are not always
good. 

The Einasto parameters
  obtained for the EMPs are in good agreement 
  with those reported for the MW halo by \citet{deason2011b} and
  \citet{sesar2011}: $ \alpha \sim  0.60, 0.45 $ and $r_2 \sim 20, 22$
  kpc, respectively.  Instead, the total accreted stellar haloes yield profiles
  slightly more concentrated (i.e. smaller $r_2$). This finding 
  suggests that the MW halo might have been formed by a significant
  contribution of accreted stars from small 
  satellites, with less than
  $20\% $ of {\it in situ } stellar mass. Under this assumption and using
  the reported observed values,
 our fitting relations  predict the DM halo of the MW to have $\alpha \sim  0.16- 0.13$ and $r_2
  \sim 13 -16 $ kpc. 
Even more, from recent measures of the
 DM density in the solar neighborhood, we can  normalize
 the simulated relation.  The resulting MW predicted profile yields an
 enclosed DM mass at $\sim 8$ kpc 
 of  $\sim 3.9 \times 10^{10}$ M$_{\odot}$ or   $\sim 6.7  \times
 10^{10}$ M$_{\odot}$, if we assume  $\rho_{\rm dm} =0.0087$ M$_{\odot}$
 pc$^{-3}$ \citep{read2014} or  $\rho_{\rm dm} =0.015$ M$_{\odot}$
 pc$^{-3}$  \citep{piffl2014}, respectively.  The agreement with the
 RAVE results by  \citet{piffl2014} are very encouraging. Detailed near future
 observations of the MW stellar halo will help us  to further  constrain
 our models and predictions.

\begin{figure*}
\hspace*{-0.8cm}\resizebox{5.0cm}{!}{\includegraphics{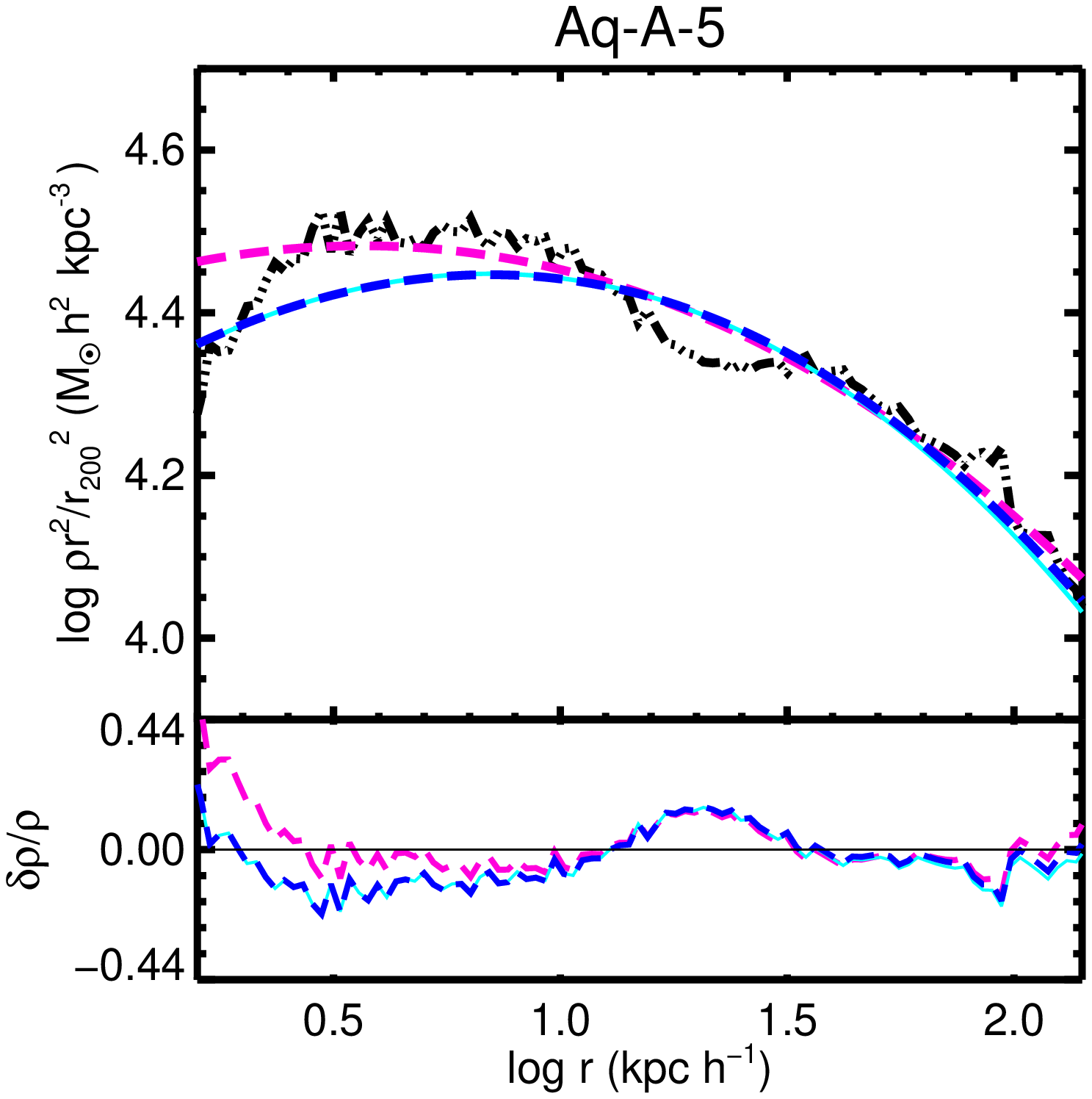}}
\hspace*{-0.8cm}\resizebox{5.0cm}{!}{\includegraphics{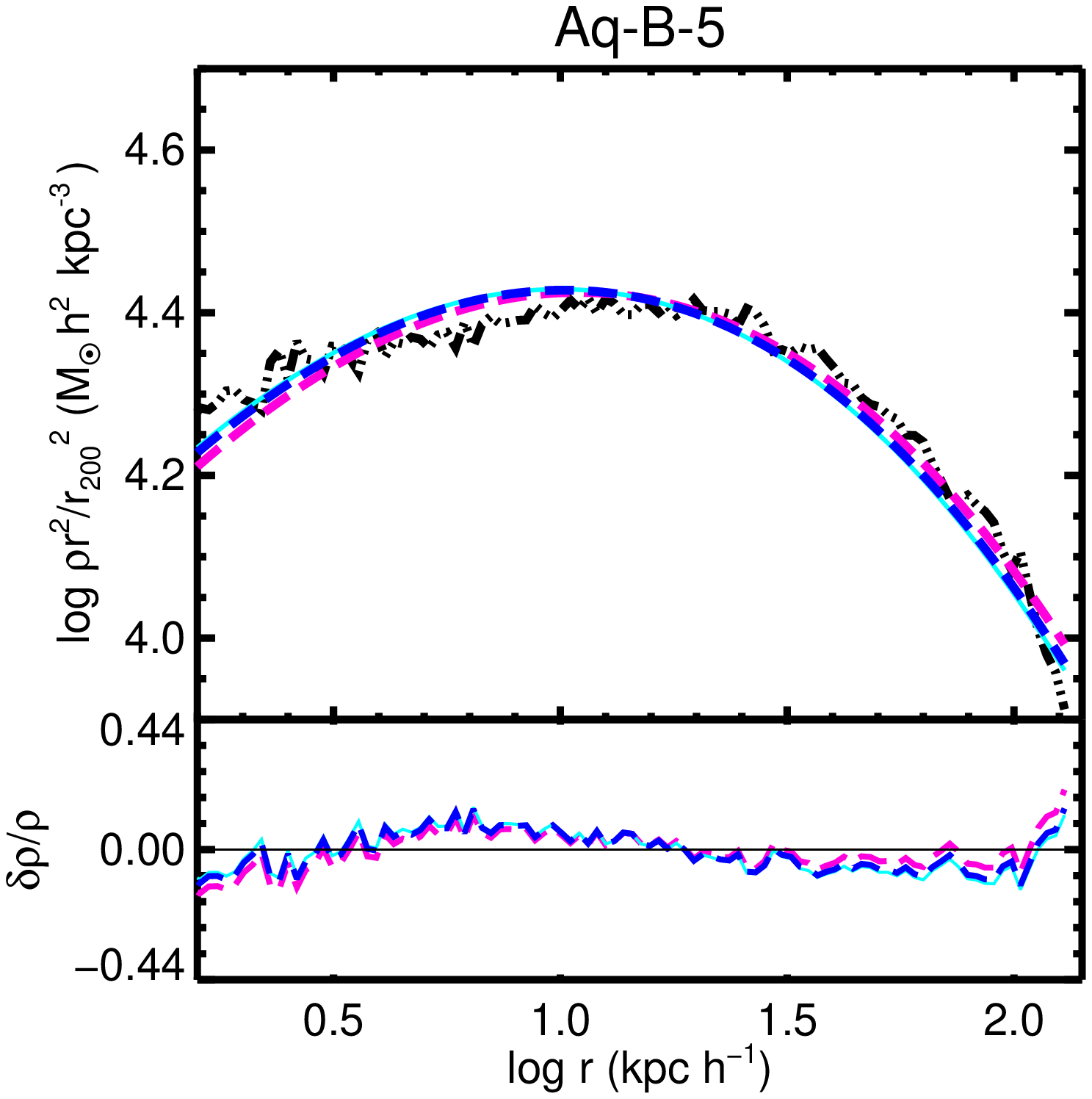}}
\hspace*{-0.8cm}\resizebox{5.0cm}{!}{\includegraphics{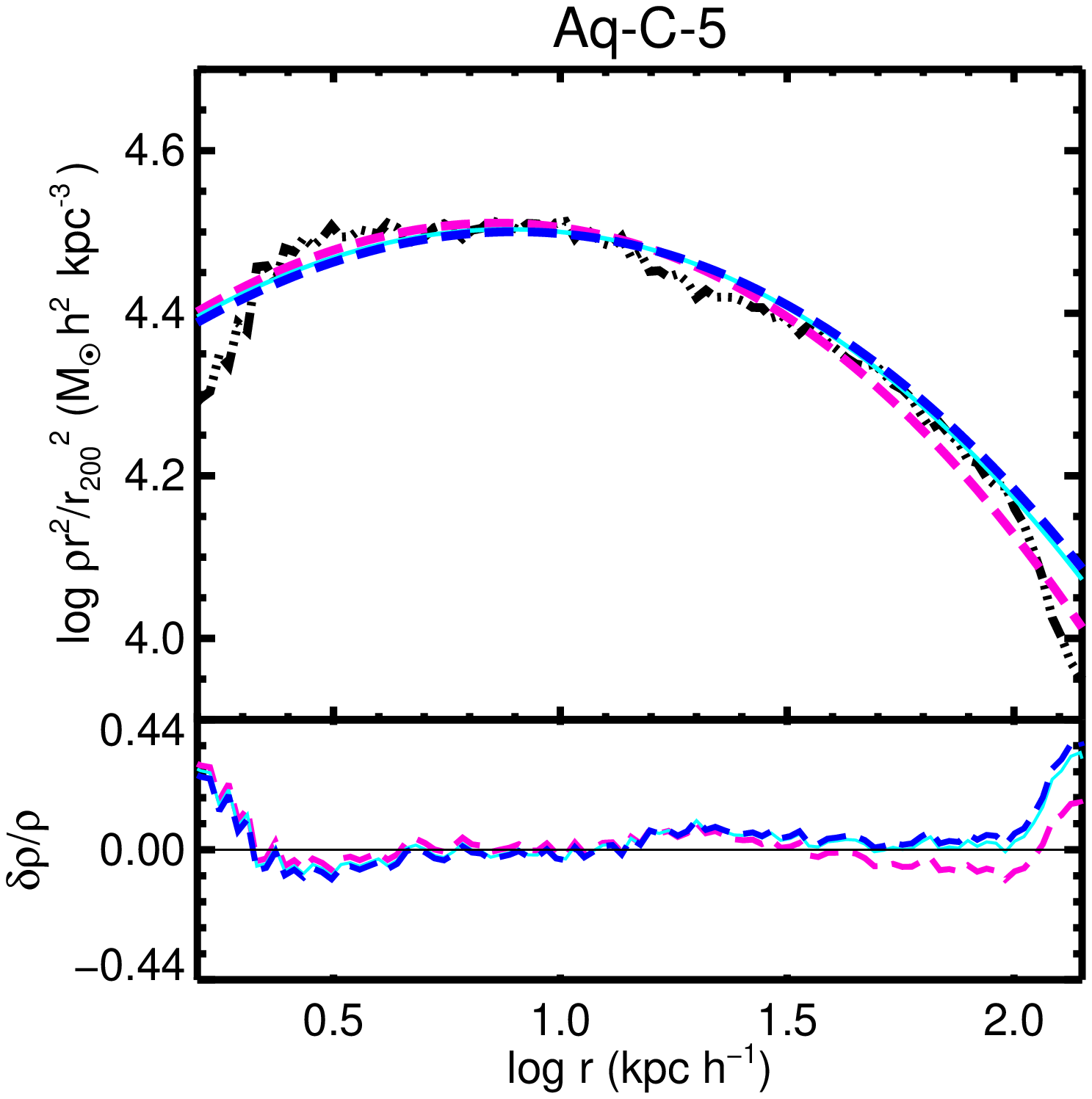}}
\hspace*{-0.8cm}\resizebox{5.0cm}{!}{\includegraphics{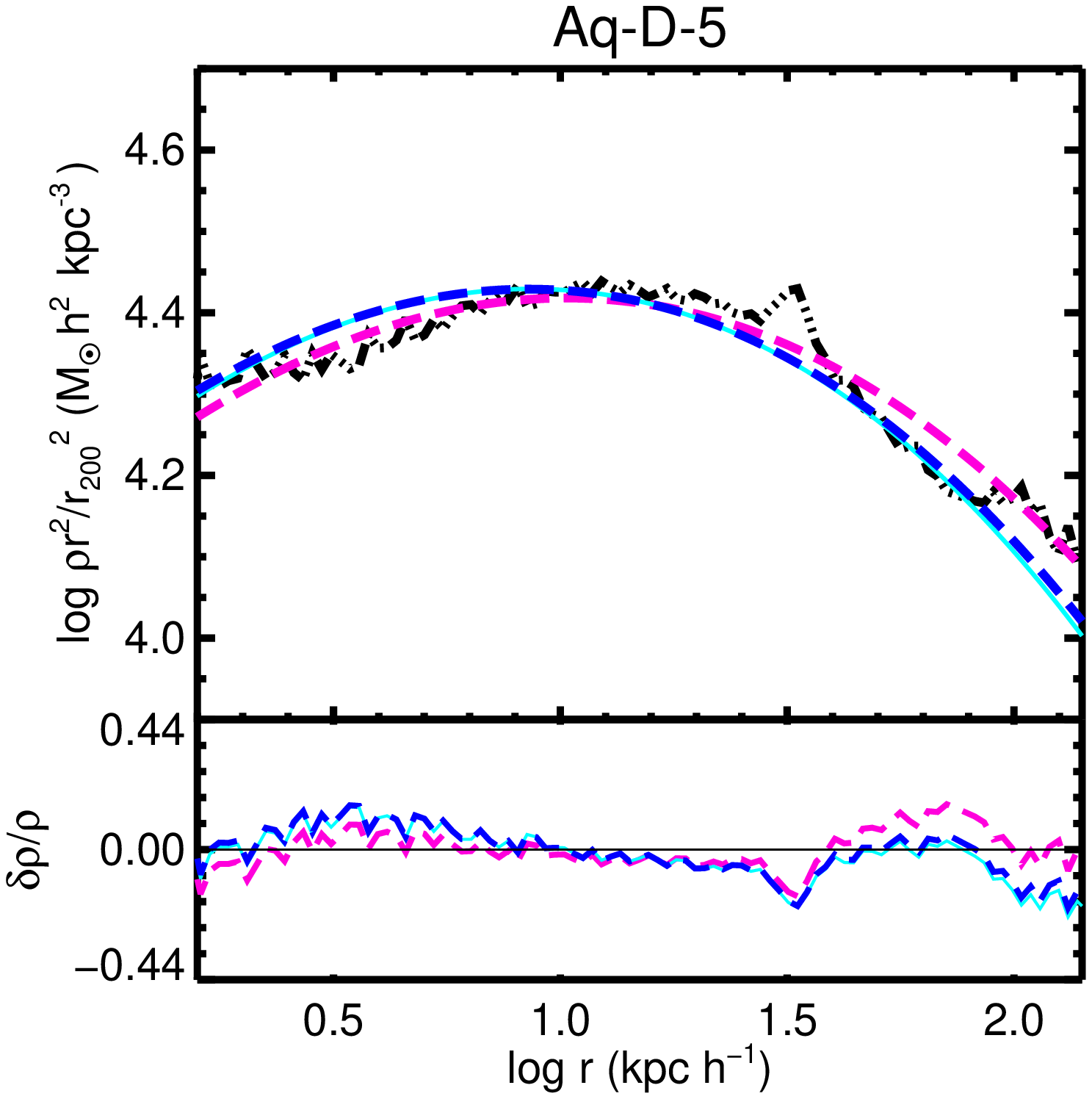}}\\
\vspace*{-0.5cm}\hspace*{-0.8cm}\resizebox{5.0cm}{!}{\includegraphics{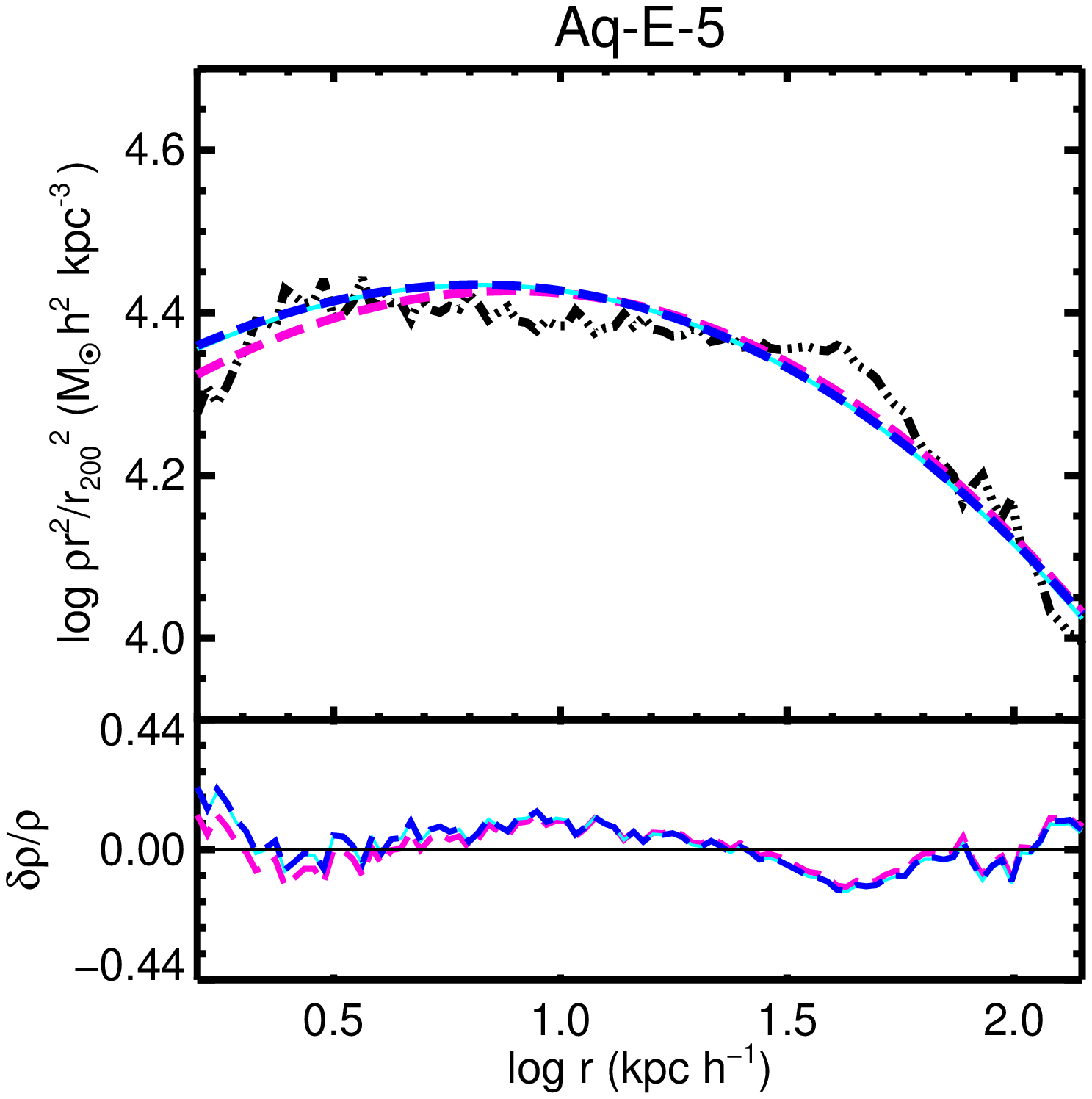}}
\hspace*{-0.8cm}\resizebox{5.0cm}{!}{\includegraphics{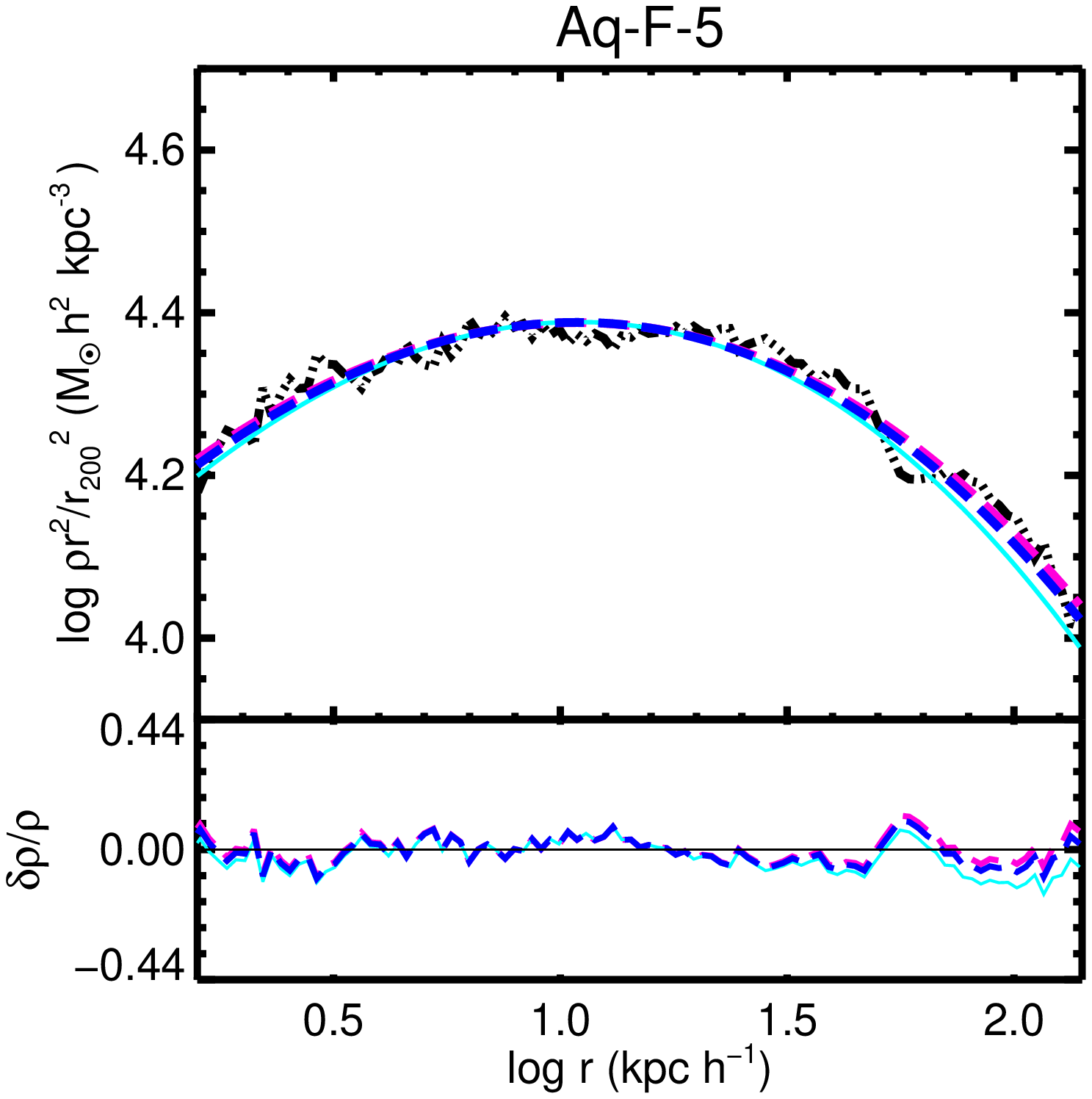}}
\hspace*{-0.8cm}\resizebox{5.0cm}{!}{\includegraphics{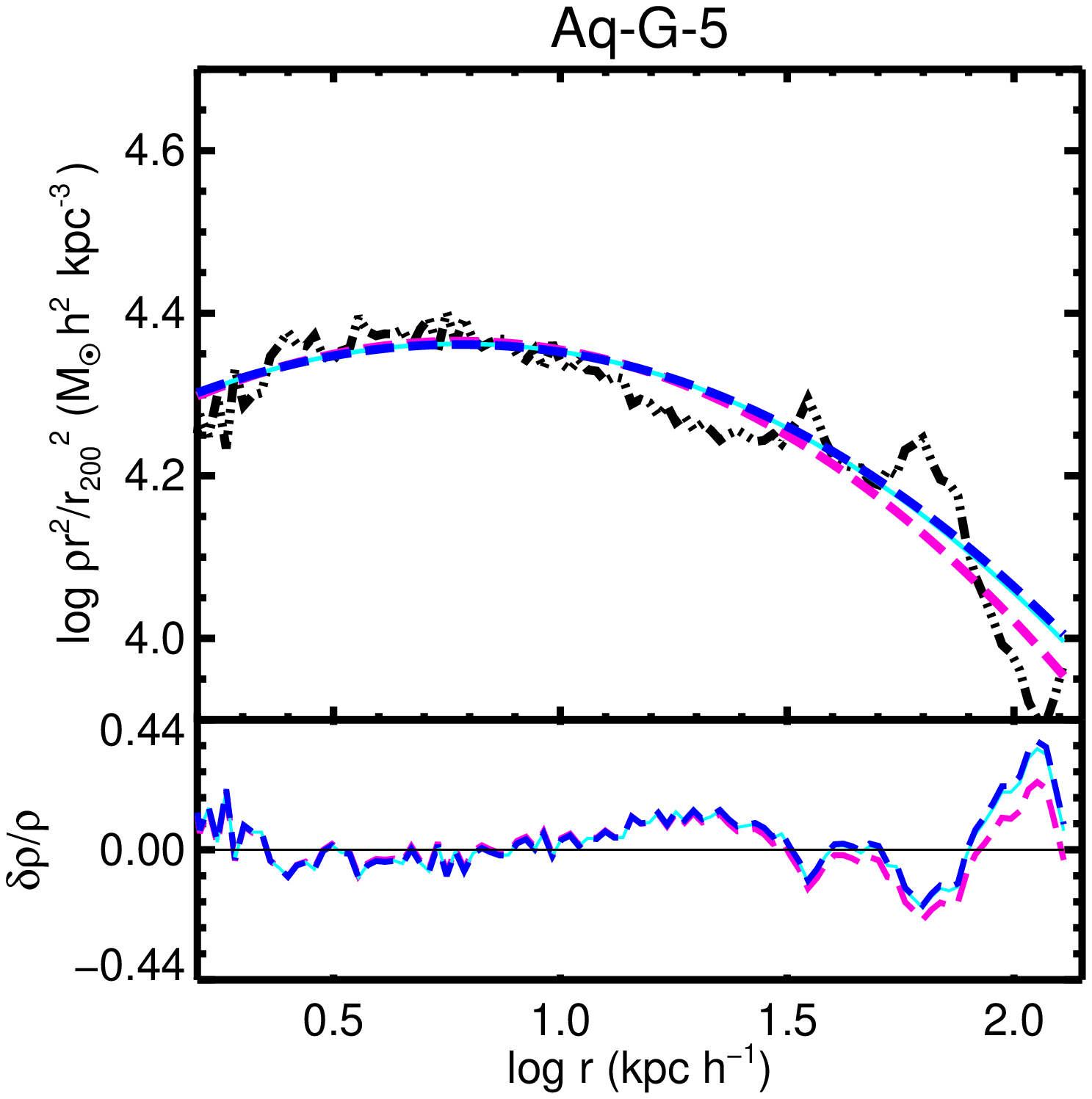}}
\hspace*{-0.8cm}\resizebox{5.0cm}{!}{\includegraphics{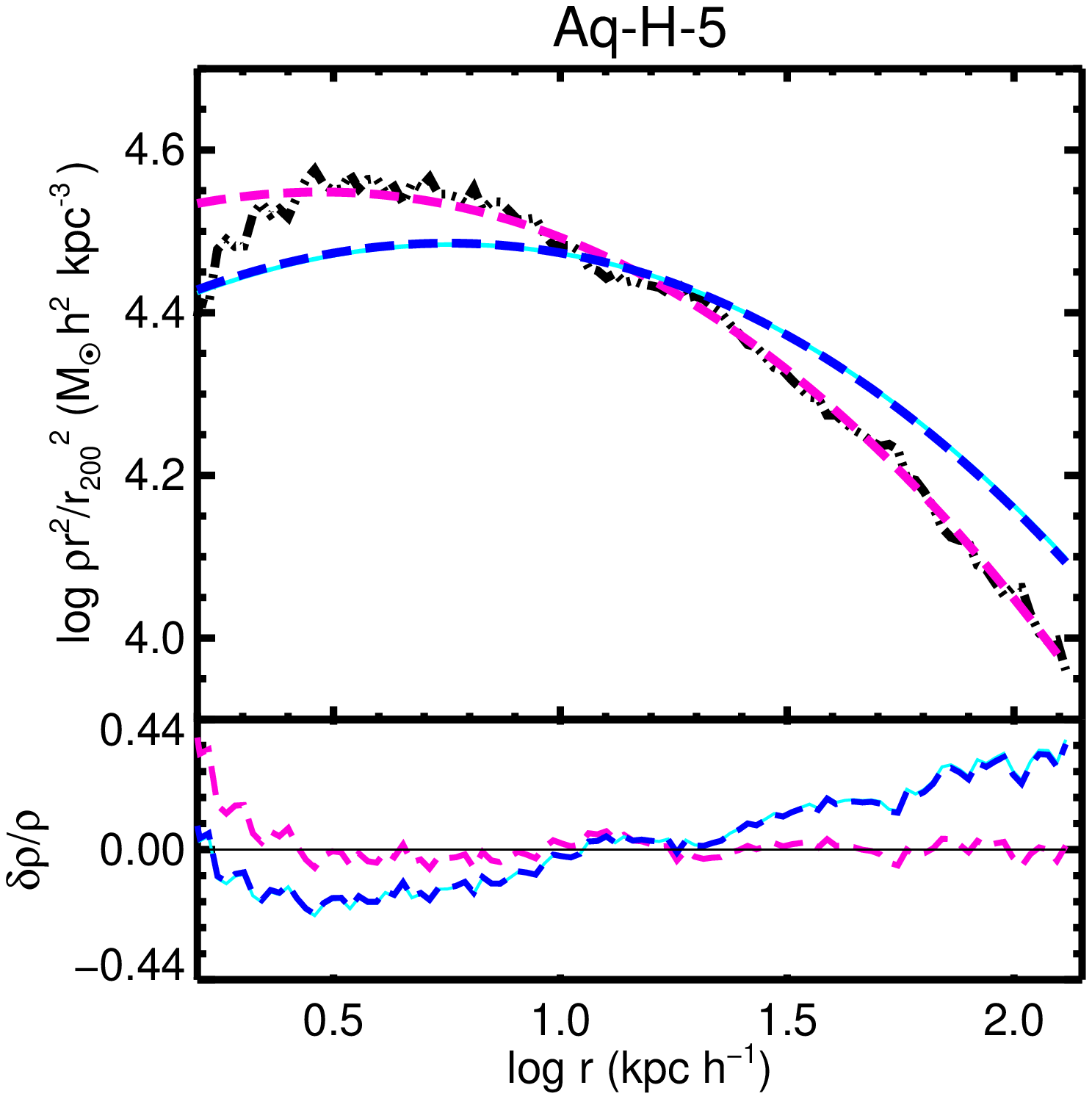}}
\vspace{-0.5cm}
\caption{ Predicted DM profiles from the fitting relations
  shown in Fig.~\ref{corre_dark} for the accreted EMPs (blue dashed
  lines), the whole EMPs (cyan solid lines), the direct
  Einasto fits to the
  DM profiles (magenta dashed
lines) and the real DM profiles from the simulations (black lines).
In the lower panels, we show the residuals of the direct Einasto fit to
the DM profiles and of those predicted
by the accreted EMP and whole EMP profiles  to the simulated DM profiles. Note that the predicted profiles  have been renormalized to reproduce
the DM profiles at $\sim r_2$ (see the text for comments). 
 }
\label{dmpro_predicted}
\end{figure*}

\section{Conclusions}

We analysed the density profiles of the stellar populations in
eight MW simulated galaxies and compared them with those of their dark
matter haloes.  We focused on  low-metallicity stars since they map the whole
potential well with a frequency increasing outwards. These stars tend to come in
low-mass satellites and hence, did not experience dissipation within
the host potential well.   

The density profiles of accreted stars  in the  stellar haloes can be well-fitted by
an Einasto profile. 
Similarly, the density profiles of accreted stars with systematically
lower [Fe/H] abundances can all be well-fitted by Einasto  profiles.
This is not the case for the total stellar populations formed by {\it in situ}
and accreted stars, since the former are much more concentrated,
producing a total density profile which is best described by a broken
power law.

The Einasto parameters obtained for the EMPs density profiles 
correlate  well with the corresponing ones  of the dark
matter profiles (even if the in situ components are included, since
they represent less than 20 per cent of the mass). The correlations get weaker as stars
with higher metallicity are included. This can be understood since more
metal-rich stars are contributed by more massive satellites and have
more possibilities to have experienced dissipation and tend to be more
centrally concentrated \citep{tissera2014}. The obtained Einasto
parameters for the EMPs are in good agreement with those reported for the MW
stellar halo, suggesting that the later might have been formed principally by the
accretion of low-mass satellites. If we adopt the observed values under
this hypothesis, 
the simulated correlations for the EMPs predict  $\alpha \sim 0.15 $ and $r_2
\sim 15$ kpc for the DM profile of the MW. Assuming observed measures
of the local DM density to normalize predicted DM profile at $\sim 8 $ kpc yields 
an enclosed DM of $ \sim 3.9-6.7 \times 10^{10}$ M$_{\odot}$, within
this radius. These estimations
are in agreement with results from the RAVE survey \citep{piffl2014}, indicating the goodness of the predicted shape in this region.

Our findings suggest low-metallicity stellar haloes might store
important information on the DM mass and the shape of  DM profiles,
helping  to constrain the cosmological model and the details of galaxy
formation models.

\section*{Acknowledgements}
We thank the anonymous referee for his/her thoughtful comments. 
This work was partially funded by  PIP 0305 (2009) and PICT Raices 959
(2011) of the Ministry of Science and Technology (Argentina) and
'Proyecto Interno' from Universidad Andr\'es Bello.
We acknowledge the support of Cosmocomp and Lacegal FP7 Marie Curie Networks.

\bibliography{TisseraLetter_revised}

\IfFileExists{\jobname.bbl}{}
{\typeout{}
\typeout{****************************************************}
\typeout{****************************************************}
\typeout{** Please run "bibtex \jobname" to optain}
\typeout{** the bibliography and then re-run LaTeX}
\typeout{** twice to fix the references!}
\typeout{****************************************************}
\typeout{****************************************************}
\typeout{}
}

\end{document}